\documentclass[a4paper,11pt]{article}
\usepackage{pos}
\usepackage{mathtools}
\usepackage{wrapfig}
\usepackage{caption}
\usepackage{subcaption}
\usepackage{siunitx}

\newlength{\bibitemsep}
\newlength{\bibparskip}
\let\oldthebibliography\thebibliography
\renewcommand\thebibliography[1]{%
  \oldthebibliography{#1}%
  \setlength{\parskip}{\bibitemsep}%
  \setlength{\itemsep}{\bibparskip}%
}

\title{First air-shower measurements with the prototype station of the IceCube surface enhancement	}
 \ShortTitle{First air-shower measurements with the prototype station of the IceCube surface enhancement}

\author{The IceCube Collaboration \\{\normalsize \normalfont(a complete list of authors can be found at the end of the proceedings)}}

\emailAdd{hrvoje.dujmovic@icecube.wisc.edu}
\emailAdd{alanc@udel.edu}
\emailAdd{marie.oehler@icecube.wisc.edu}

\abstract{IceTop, the surface array of the IceCube Neutrino Observatory, consists of 162 ice-Cherenkov tanks distributed over an area of 1km$^2$. Besides being used as a veto for the in-ice neutrino detector, IceTop is a powerful cosmic-ray detector. In the upcoming years, the capabilities of the IceTop array will be enhanced by augmenting the existing ice-Cherenkov tanks with an array of elevated scintillator panels and radio antennas. Combining the data obtained from the different detectors will improve the reconstruction of cosmic-ray energy and primary mass while reducing the energy threshold and increasing the aperture of the array. In January 2020, a prototype station consisting of 8 scintillation detectors and 3 antennas was deployed at the IceTop site. The prototype detectors are connected to one data-acquisition system and the readout of the radio antennas is triggered using the signals from the scintillators. This allows us to regularly observe secondary air shower particles hitting the scintillators, as well as the radio emission of high-energy air showers. In this contribution, we will discuss the results obtained from the prototype station in the past year, present the first cosmic-ray air showers measured with this prototype station, and show how the observations with the different detector types complement each other.

\vspace{4mm}
{\bfseries Corresponding authors:}
Hrvoje Dujmovi\'c$^{1*}$, Alan Coleman$^{2}$, Marie Oehler$^{1}$\\
{$^{1}$ \itshape Karlsruhe Institute of Technology, Institute for Astroparticle Physics, D-76021 Karlsruhe, Germany}\\
{$^{2}$ \itshape Bartol Research Institute and Dept. of Physics and Astronomy, University of Delaware, Newark, DE 19716, USA}\\[4mm]
$^*$ Presenter

\FullConference{37$^{\rm{th}}$ International Cosmic Ray Conference (ICRC 2021)\\
		July 12th -- 23rd, 2021\\
		Online -- Berlin, Germany}

}

\begin{document}
\maketitle

\section{Introduction}\label{sec:intro}
The IceCube Neutrino Observatory is a multi-purpose experiment located at the geographic South
Pole. The cubic-kilometre-sized neutrino detector is placed inside the ice at depths between 1450\,m to 2450\,m. IceTop, consisting of ice-Cerenkov tanks on the surface, acts as a veto against atmospheric muons and is also used as a powerful cosmic-ray detector~\cite{IC_instrumentation}.

Each year, additional snow is deposited on top of the IceTop tanks. Over time, this has lead to a significant increase in the IceTop energy threshold. Additionally, the uneven snow coverage and the uncertainty in the particle interactions inside the snow have introduced additional systematic uncertainties to the cosmic-ray air-shower measurements~\cite{snow_attenuation}. To counteract these problems and open up new possibilities for cosmic-ray measurements, an IceTop-enhancement will be deployed in the upcoming years. The array will comprise 32 stations, each consisting of eight scintillator panels and three radio antennas~\cite{tom2017icrc, frank2019icrc, agn2019icrc}. 
In order to test the design and performance of these stations, a prototype station has been deployed at the South Pole in January 2020. The station contains eight scintillator panels with a sensitive area of 1.5\,m$^2$ each~\cite{matt2019icrc}. The three antennas are SKALA-2 antennas which can independently measure the two polarisations in the frequencies between 70-350 MHz. The layout of the station can be seen in Figure~\ref{fig:event_display}, more details on the station hardware can also be found in~\cite{marie2021icrc}.
This station builds on experiences gained from previous prototypes~\cite{tom2017icrc, matt2019icrc, max2019icrc}. However, these earlier prototypes had major hardware differences compared to the current station. The 2020 prototype is nearly identical to the stations that will be deployed for the full enhancement array. Some minor improvements are discussed in~\cite{marie2021icrc}.

After an initial commissioning phase, air-shower data has been taken with the prototype station since 2020-10-22 and the data until 2021-01-27 is studied in this work. 

\section{Data acquisition and processing}\label{sec:processing}
For air-shower data-taking, the detector is configured in a way that each of the eight scintillator panels records a hit if the recorded charge is above 3612 ADC counts in the high-gain channel. In the future, a temperature-corrected threshold is planned to be used with a threshold of 0.5 MIP. However, due to varying environmental conditions, the current threshold value can vary between zero and a few MIPs.
If a hit is recorded, the total charge deposited in the panel and the time of the hit are saved. 

If at least 6 of the scintillator panels record a hit within \SI{1}{\micro\second}, the readout of the radio antennas is triggered. For each of the two polarisations in all three antennas, 1024 samples are recorded with a sampling frequency of 1\,GHz. In addition to the digitised waveforms, the timing of the trigger is also recorded. 

\subsection{Radio data}\label{sec:processing_radio}
For the radio data, each trace is split into four and then sampled in parallel four times. The median of the four traces is taken as the measured signal. This procedure helps to suppress noise spikes that can otherwise be observed in individual traces. These spikes do not significantly contribute to the overall noise power, but can sometimes be confused for radio signals from air-showers. The exact origin of these noise spikes is still under investigation, however they do not seem to be correlated between the four copies of the trace and are thus most likely caused by issues in the data digitisation. Taking the median of the four traces ensures any such noise spikes are filtered out.

In the next step a background-weighted frequency filtering is applied. The idea here is to suppress frequencies which are known to have significant RFI contributions and promote frequencies with low RFI. The median spectrum obtained from calibration measurements in the same time period is taken as the background spectrum. More details on this method can be found in~\cite{alan2021icrc}.
For each antenna $i$ we then use the weighted waveforms $W_{ij}$ and define: $F_i(t) = \sum_{j}|H(W_{ij}(t))|^2$ as a proxy for the energy fluence in the antenna. Here, $H$ denotes the Hilbert transformation and the sum is performed over the two polarisations, $j$. The signal time is defined as the position of the maximum of $F_i$ and the signal-to-noise ratio (SNR) is defined as the ratio, $\text{SNR}_i \coloneqq \text{max}_t(F_i)/\text{median}_t(F_i)$.
Showers with radio signals are selected based on their SNR$_i$ value.

\subsection{Scintillator data}\label{sec:processing_scint}
The hits from the scintillators are received in 8 independent data streams, one from each panel. These are then merged into events by using a \SI{1}{\micro\second} sliding window. In order to make the directional reconstruction possible, events with less then 3 panel hits are rejected.

\begin{figure}[ht]
\centering
\begin{minipage}{.49\textwidth}
  \centering
  	\includegraphics[width=.91\linewidth]{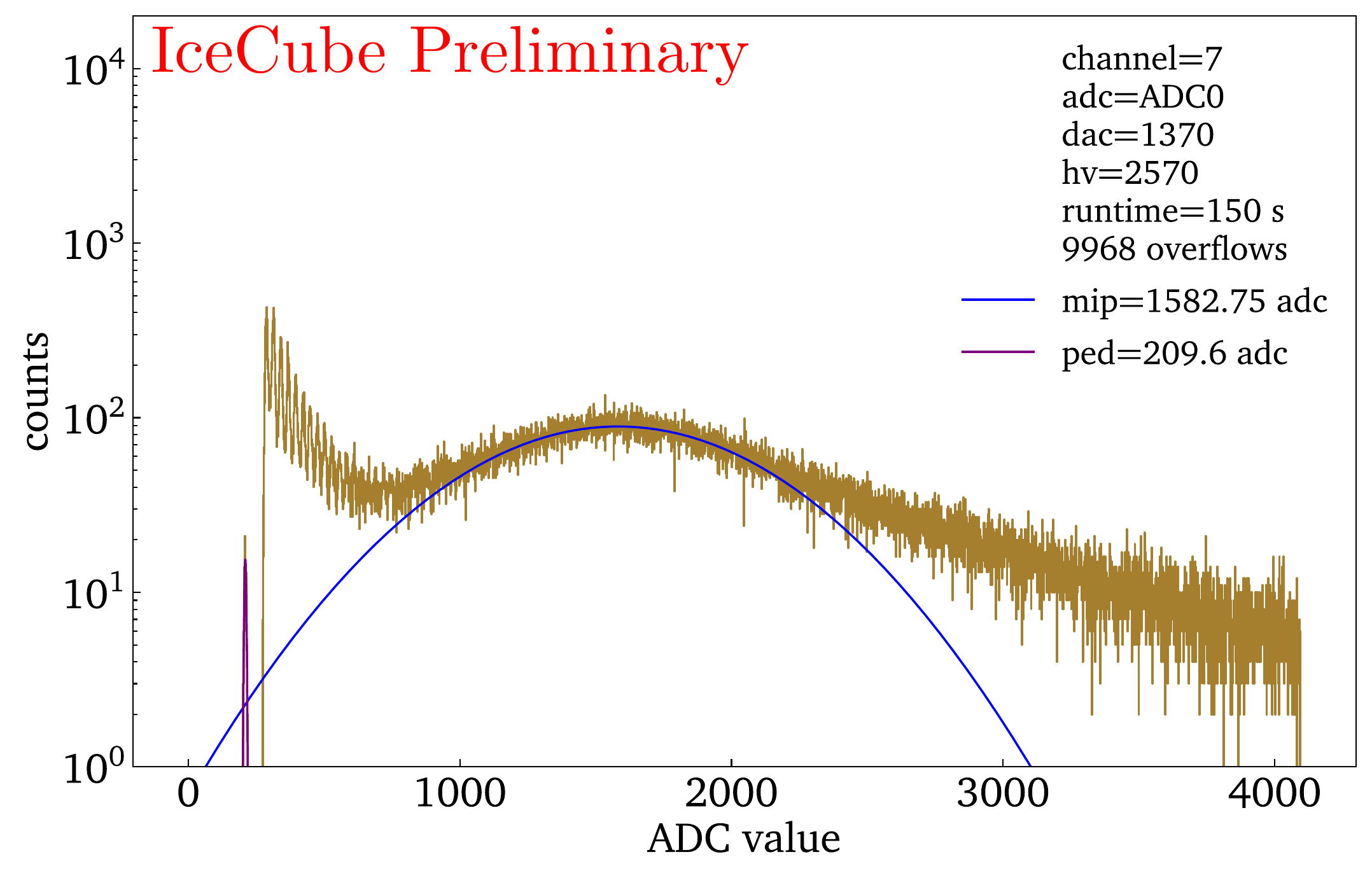}
	\captionof{figure}{Scintillator calibration data from a single run. The charge of each hit is recorded and histogrammed. The MIP-peak is then fitted with a Gaussian function.}
    \label{fig:MIPfit}
\end{minipage}
\hfill
\begin{minipage}{.49\textwidth}
  \centering
	\includegraphics[width=.8\linewidth]{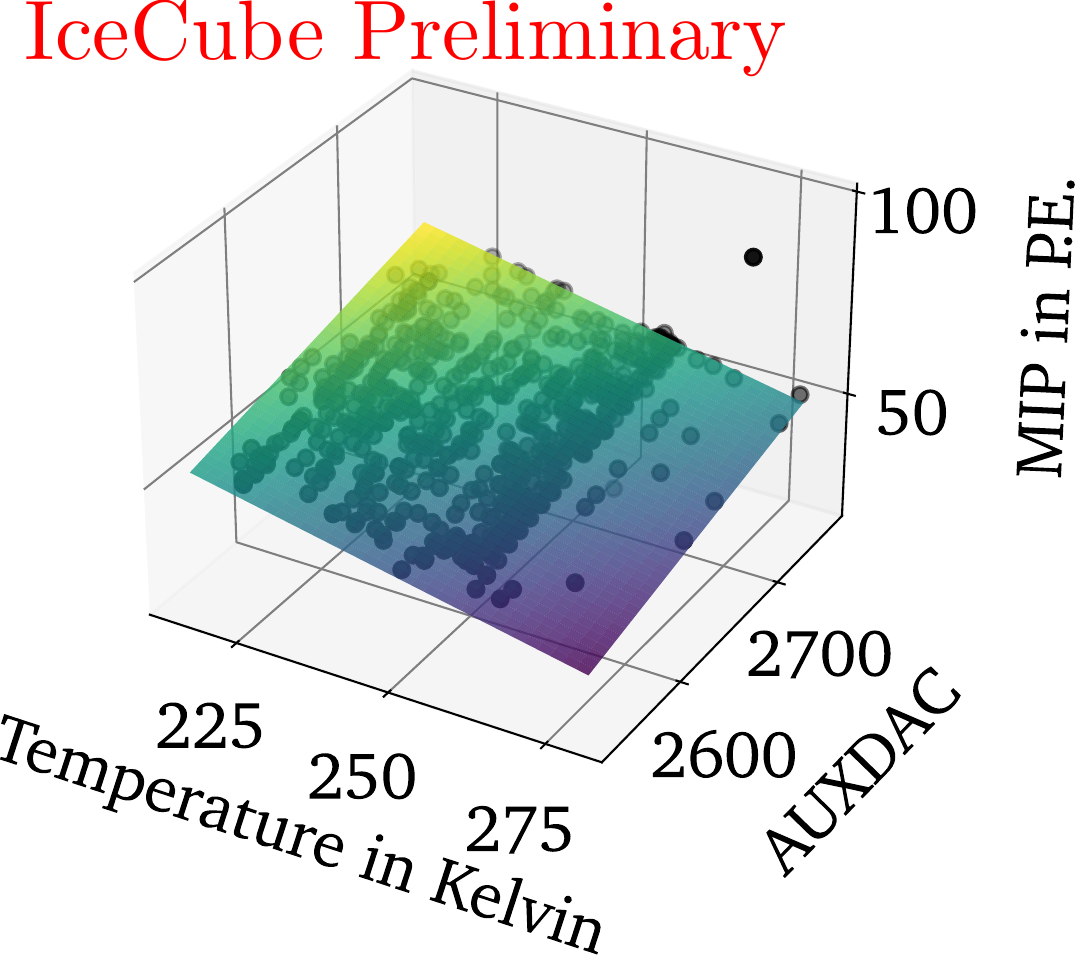}
	\captionof{figure}{Each point represents a single calibration run performed at a specific temperature and using a specific SiPM bias voltage setting (AUXDAC). The light yield is plotted in the z-axis and fitted by the a plane, see Equation~\ref{eq1}.}
    \label{fig:THVfit}
\end{minipage}
\end{figure}

The charge deposited in the scintillators is read out in ADC units. To use these charges for reconstruction, they need to be compared to the average charge from a single MIP (minimum ionizing particle). An additional complication arises from the fact that the ADC to MIP conversion factor has a strong temperature-dependence. Additionally, the SiPM (silicon photo-multiplier) gain depends on the applied bias voltage, which can also be adjusted during operation. Ideally, the data acquisition system can be programmed to monitor the temperature and adjust the bias voltage in order to keep the gain stable~\cite{matt2019icrc}. However, this functionality has not been implemented in the current system, so these gain changes have to be corrected for a posteriori. For this purpose, we perform calibration runs with the detector threshold set to a very low value of a few PE. There is no time information of the hits being recorded during these calibration runs. Additionally, during these runs, a periodic soft-trigger is sent to the DAQ and the charge corresponding to a baseline trace is recorded. The histogrammed data from one such run can be seen in Figure~\ref{fig:MIPfit}. The soft-triggered charge values from a baseline is first subtracted from all the other charge values. The local MIP maximum is found and a fit interval is set to be between the values where the count falls off to 0.607 times the local MIP maximum. A Gaussian function is then fitted to the distribution in the interval. The mean of the Gaussian is then taken as the MIP-gain. These MIP-gain values are compared to the PE-gains calculated in~\cite{marie2021icrc} to get the MIP light yield (LY) values. 

The calibration runs are performed with varying SiPM bias voltage settings and on days with different ambient temperatures. The LY values from these various calibration runs can be seen in Figure~\ref{fig:THVfit}. Empirically, the data seems to be well-described by the bilinear function, see Equation~\ref{eq1}. The voltage here is given in the arbitrary AUXDAC units. 
For each data-taking run, the temperature and the SiPM voltage are also logged. The charges in ADC units can then be transformed into a photoelectron-number as described in~\cite{marie2021icrc} and then to a number of MIPs. 

\begin{equation}\label{eq1}
\begin{split}
\text{LY}&=C_T \cdot T + C_V \cdot V + C_0 \\
C_T &= -3.21\cdot 10^{-1}\ K^{-1},\ C_V = 3.70\cdot 10^{-3},\ C_0 = 8.57\cdot 10^{-2}
\end{split}
\end{equation}

\subsection{Coincident events}\label{sec:processing_coincidences}
\begin{wrapfigure}{R}{0.55\textwidth}
\centering
	\includegraphics[width=.52\textwidth]{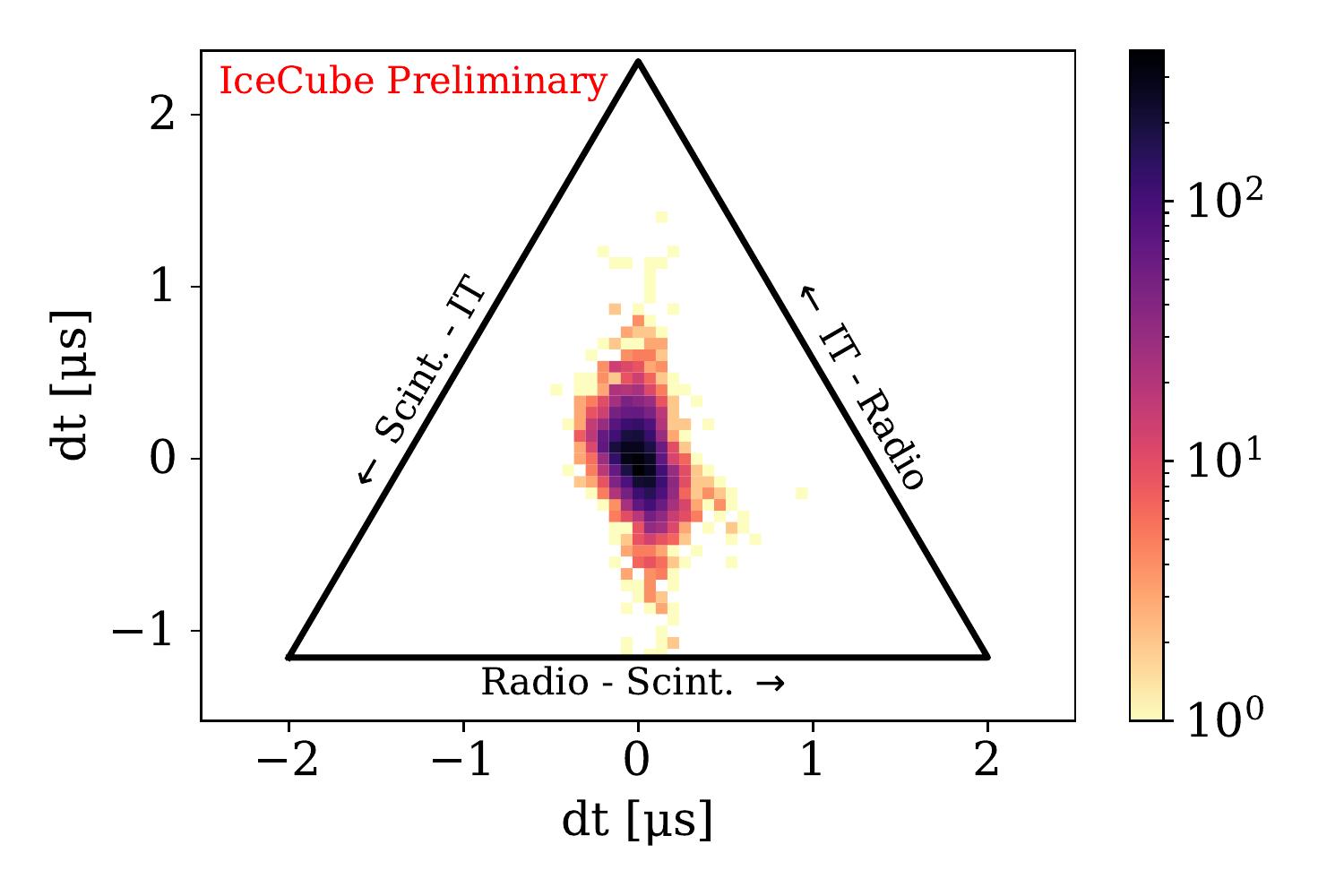}
	\protect\caption{Differences in the event timing between radio, scintillators and IceTop. Events inside the triangle are considered to be in coincidence. The colour scale represent the number of events/bin.}
    \label{fig:time_triangle}
\end{wrapfigure}

At this stage, the surface data is received in three different streams; scintillator, radio and IceCube (including the IceTop data) streams. These data streams are merged based on event timestamps. For radio data, the trigger times of events are used. For scintillators, the time of the first panel hit is used. For IceTop data, a standard IceTop reconstruction is performed and the reconstructed time of the shower hitting the surface is used. If a radio, a scintillator, and an IceTop event are found within a \SI{2}{\micro\second} window, the events are merged into a single coincidence event. An illustration of the timing differences can be seen in Figure~\ref{fig:time_triangle}. The time window of \SI{2}{\micro\second} seems to be large enough to allow for nearly all physical events to pass the cut while keeping the rate of accidental coincidences low. 

\section{Air shower reconstruction}\label{sec:reconstruction}
For events where scintillator, radio and IceTop data are recorded, an event reconstruction is performed on the three components and the results are compared to check for consistency between the detector components.

For IceTop data, the events are reconstructed using the standard IceTop event reconstruction, as can be found in~\cite{IT_energy_reco}. This reconstruction provides a shower direction, core position and S$_{125}$, an energy-proxy parameter.

\newpage
For scintillator signals, the shower core is estimated as the weighted centre of the scintillator positions, where the weights are given by the charge deposited in the panel. The shower direction is then reconstructed assuming a plane shower front and minimising the mean square of the residuals between the expected signal times and the measured signal times.

\begin{figure}[ht]
    \centering
	\includegraphics[width=.98\textwidth]{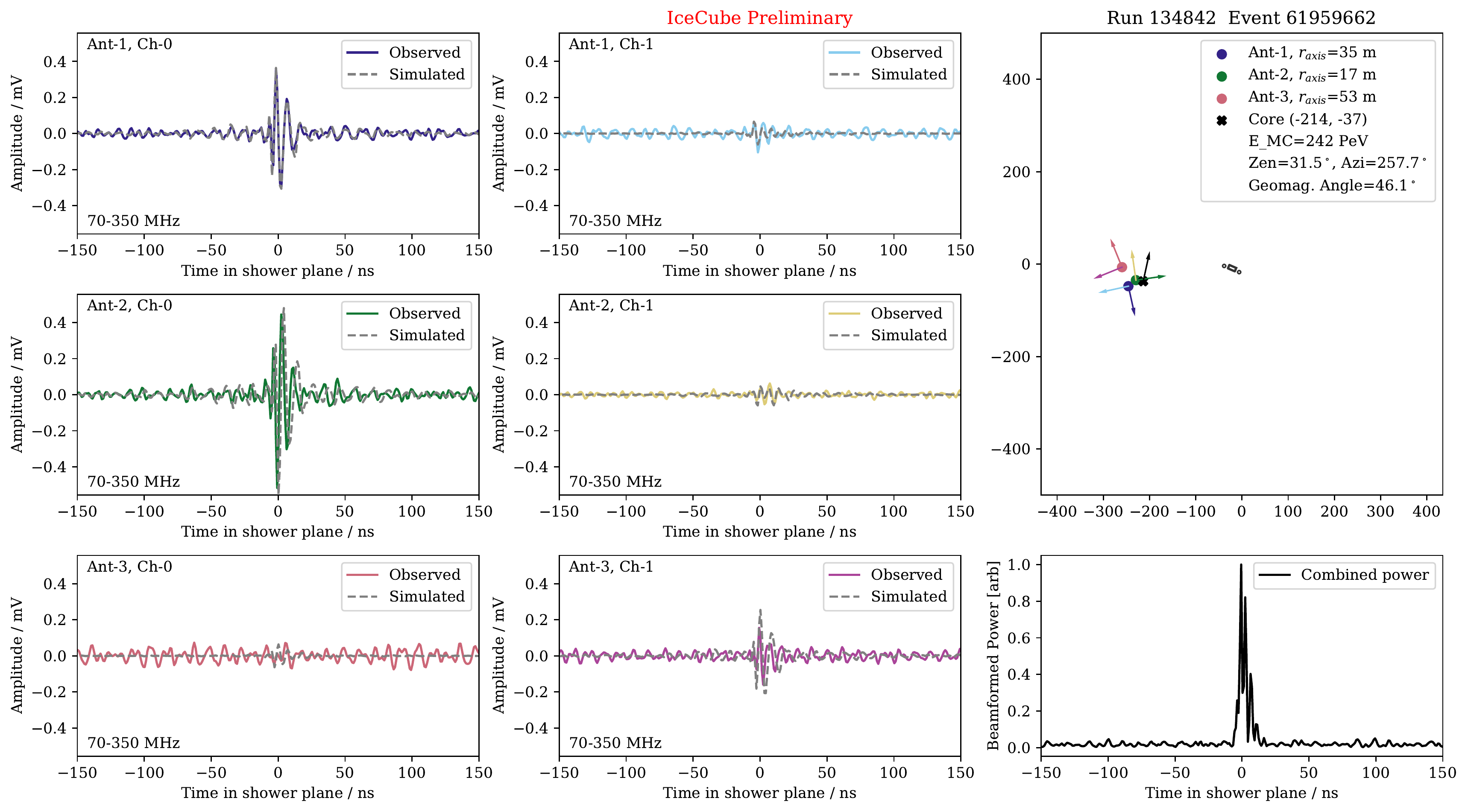}
	\protect\caption{For an identified radio event, a simulation is performed using the shower parameters reconstructed with IceTop. The left and center columns show the observed waveforms after frequency filtering (solid) and the simulated waveforms (dashed) for the six channels. No additional noise has been added to the simulation. In the top right, one can see a map of the detector with the antenna positions and orientations of the two polarisation arms. In the bottom right one can see the combined radiation power of the beam-formed waveforms from data under the assumption that the shower direction is as reconstructed by IceTop.}
    \label{fig:mc_wf}
\end{figure}

Since the threshold for the scintillator triggers is much lower than the threshold for detection of radio showers~\cite{agn2019icrc, alan2021icrc}, most of the triggered events do not have identifiable radio pulses in them. Thus additional selection cuts are applied to the radio events before performing the reconstructions. 
Radio events are discarded if the SNR in any of the three antennas is less than 7.0. Additionally, events where the signal times of all three antennas are within 5\,ns are rejected. Lastly, events are rejected where the peak time separation is larger then the physical distance between the antennas. These cuts have been developed on software-triggered background data. The second cut introduces a suppression for events arriving from a specific direction, however it helps with suppressing electronic noise spikes which show a high degree of correlation in time.
After the additional cuts have been applied, the event directions are reconstructed assuming a plane shower front in the same way as it has been done for the scintillator data. 
Additionally, for the radio showers that pass those cuts, radio simulations are performed assuming a shower core position, direction and energy as reconstructed by IceTop. The radio waveforms from the simulations are then visually compared to the observed ones and the overall pulse shapes and amplitudes seem to match. At this stage, this offers no additional quantitative information for the shower reconstruction. It does, however, offer an important qualitative cross-check. An example of such a data-simulation comparison can be seen in Figure~\ref{fig:mc_wf}. 

\begin{figure}[t]
    \centering
	\includegraphics[width=.85\textwidth]{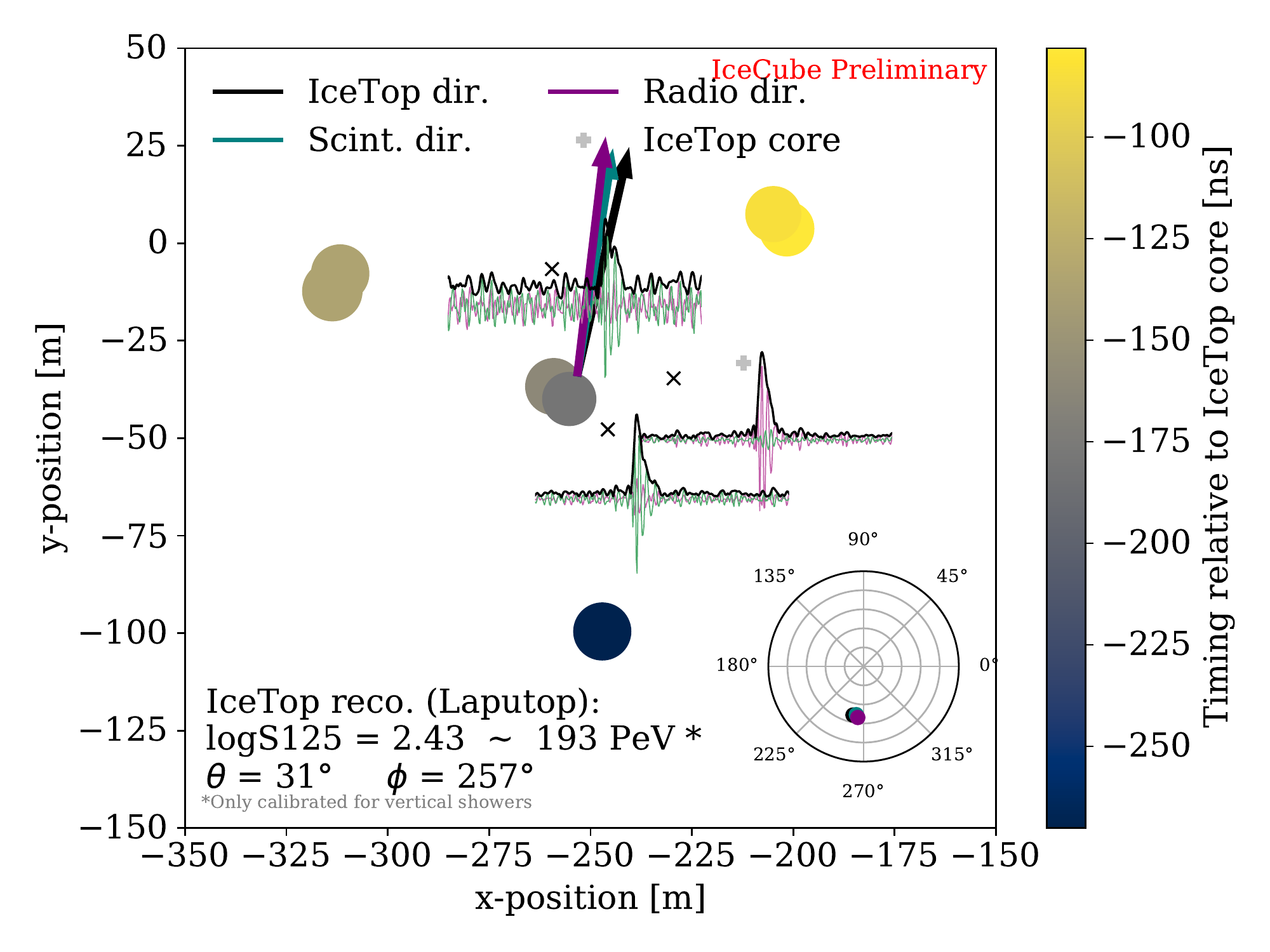}
	\protect\caption{An example of an air shower event seen in all three detectors. Coloured circles show the locations of the scintillator panels that saw a hit. The size of the circles indicates the deposited charge, the colour indicates the timing relative to IceTop reconstructed core time. The IceTop reconstructed core position is indicated by the gray +. The three x indicate the positions of the radio antennas. For each of the antennas, the filtered waveforms of the two polarisations and the sum of their envelopes is shown. The three arrows indicate the reconstructed directions from the three detectors projected onto the x-y plane. A small skymap with the arrival directions is shown in bottom right and the IceTop reconstruction parameters are shown in the bottom left. This is the same even as has been seen in Figure~\ref{fig:mc_wf}.}
    \label{fig:event_display}
\end{figure}

In Figure~\ref{fig:event_display}, the same event as in Figure~\ref{fig:mc_wf} is shown with all the surface array data and the three reconstructions.

\section{Reconstruction performance}\label{sec:distrubtions}
\begin{figure}[ht]
\centering
\begin{subfigure}{.45\textwidth}
  \centering
	\includegraphics[width=1\textwidth]{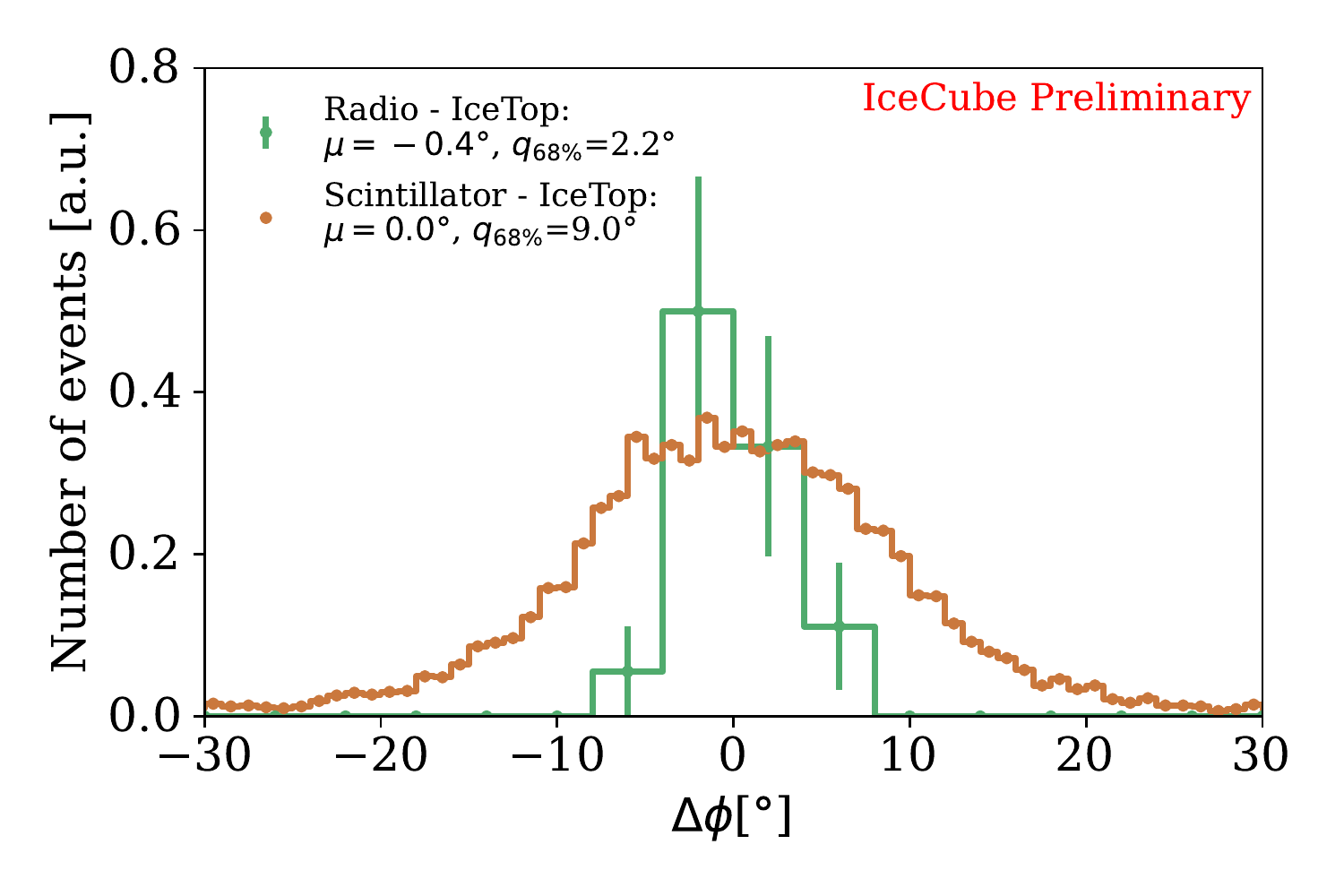}
	\protect\caption{Azimuth differences}
    \label{fig:delta_phi}
\end{subfigure}%
\begin{subfigure}{.45\textwidth}
  \centering
	\includegraphics[width=1\textwidth]{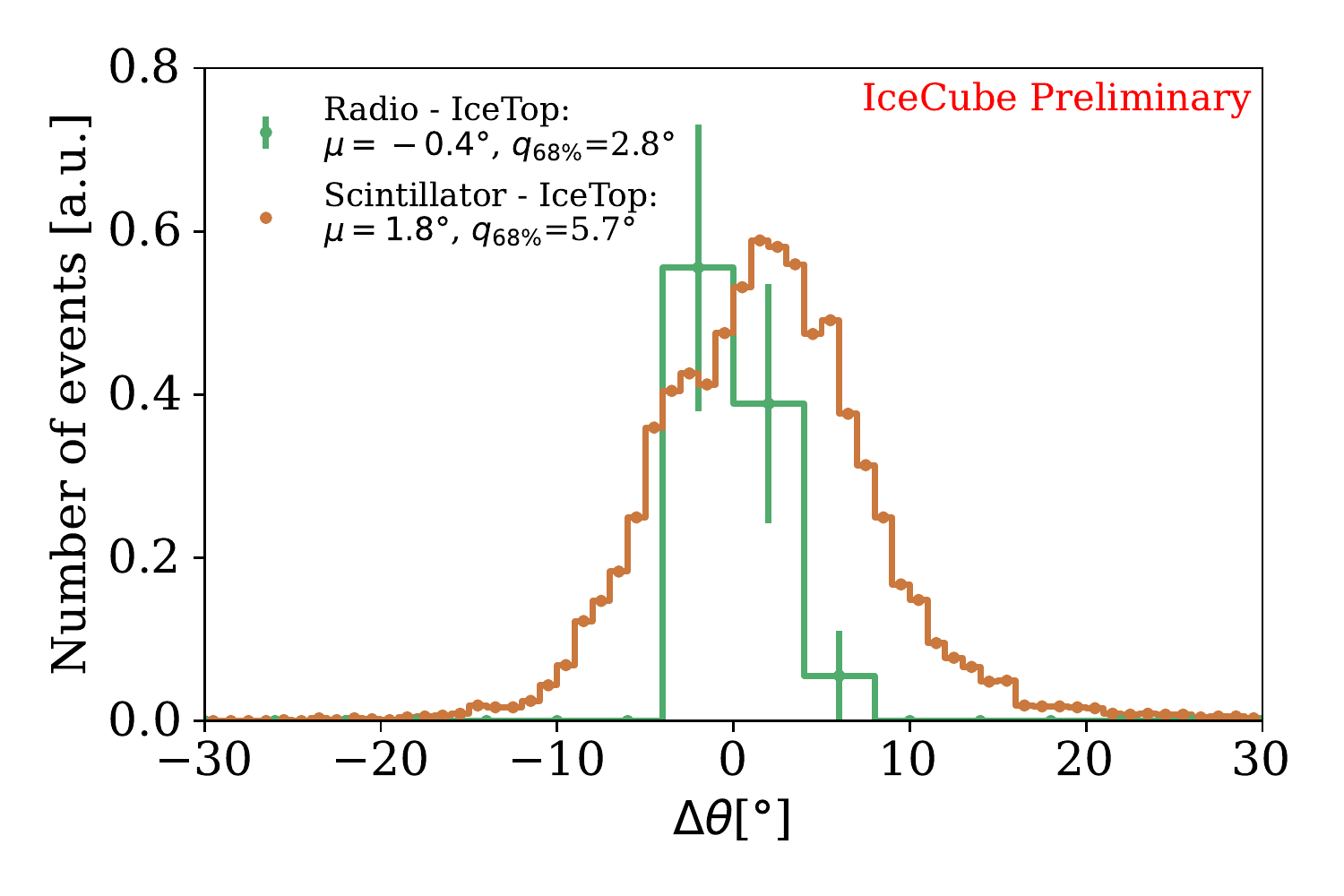}
	\protect\caption{Zenith differences}
    \label{fig:delta_zeinth}
\end{subfigure}
\caption{Differences in the directional reconstructions between the three different detector components. The error bars represent the statistical 1~$\sigma$ uncertainties. For the differences between radio and IceTop, and scintillators and IceTop, the median and the width of the central 68-percentile are calculated.}
\label{fig:delta_directions}
\end{figure}

The reconstructed shower directions are compared to each other in  Figure~\ref{fig:delta_directions}. Since IceTop is a well-established detector, the directional reconstructions from the other detector components are compared to IceTop. For the radio reconstruction comparison, only showers passing the radio cuts are used, while for scintillator reconstruction all showers are used. Overall, the reconstructed directions seems to match quite well, although the angular resolution is somewhat limited by the small footprint of the prototype station. 

\begin{figure}[ht]
\centering
\begin{subfigure}{.45\textwidth}
  \centering
	\includegraphics[width=1\textwidth]{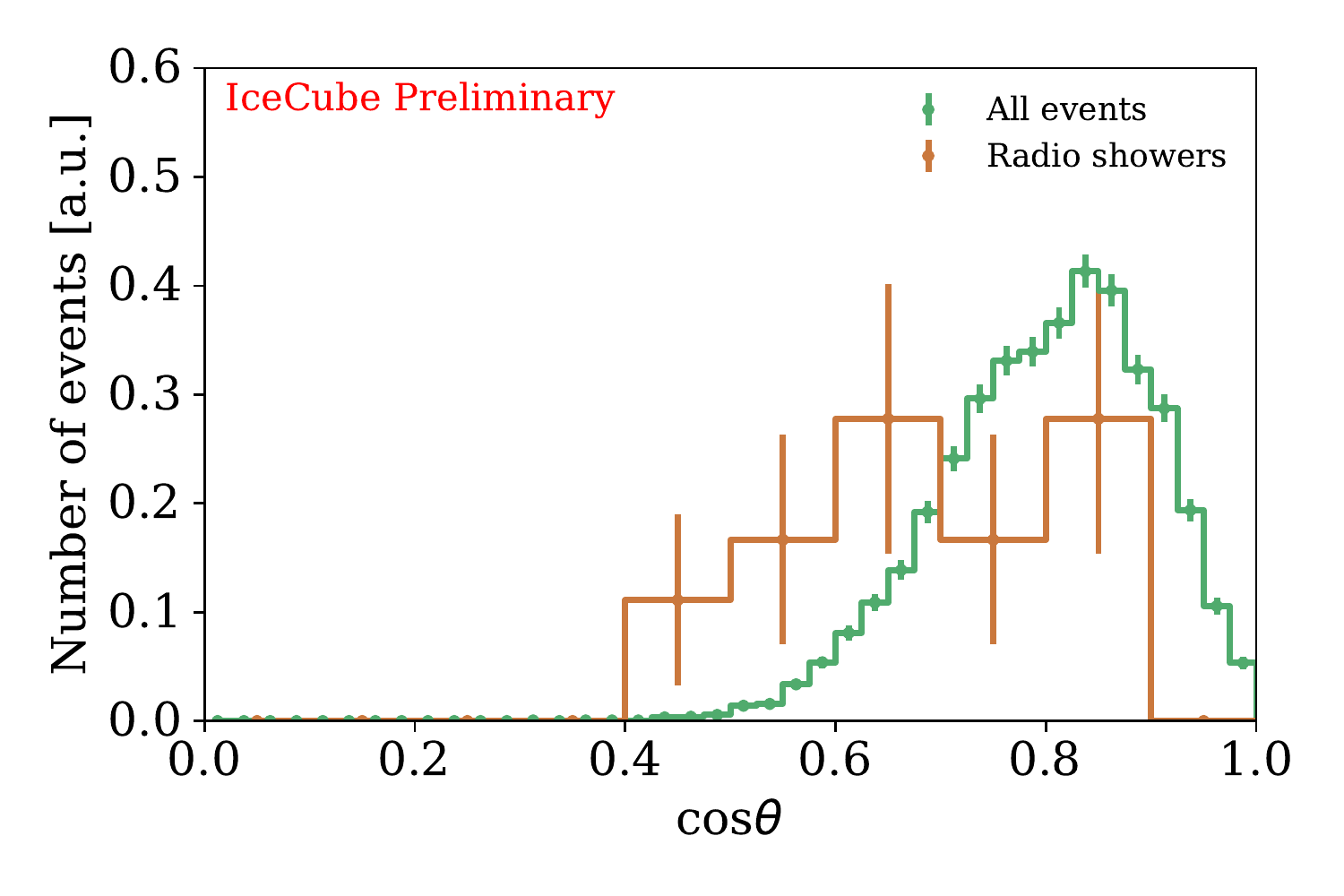}
	\protect\caption{Zenith distribution}
    \label{fig:zenith_dist}
\end{subfigure}%
\begin{subfigure}{.45\textwidth}
  \centering
	\includegraphics[width=1\textwidth]{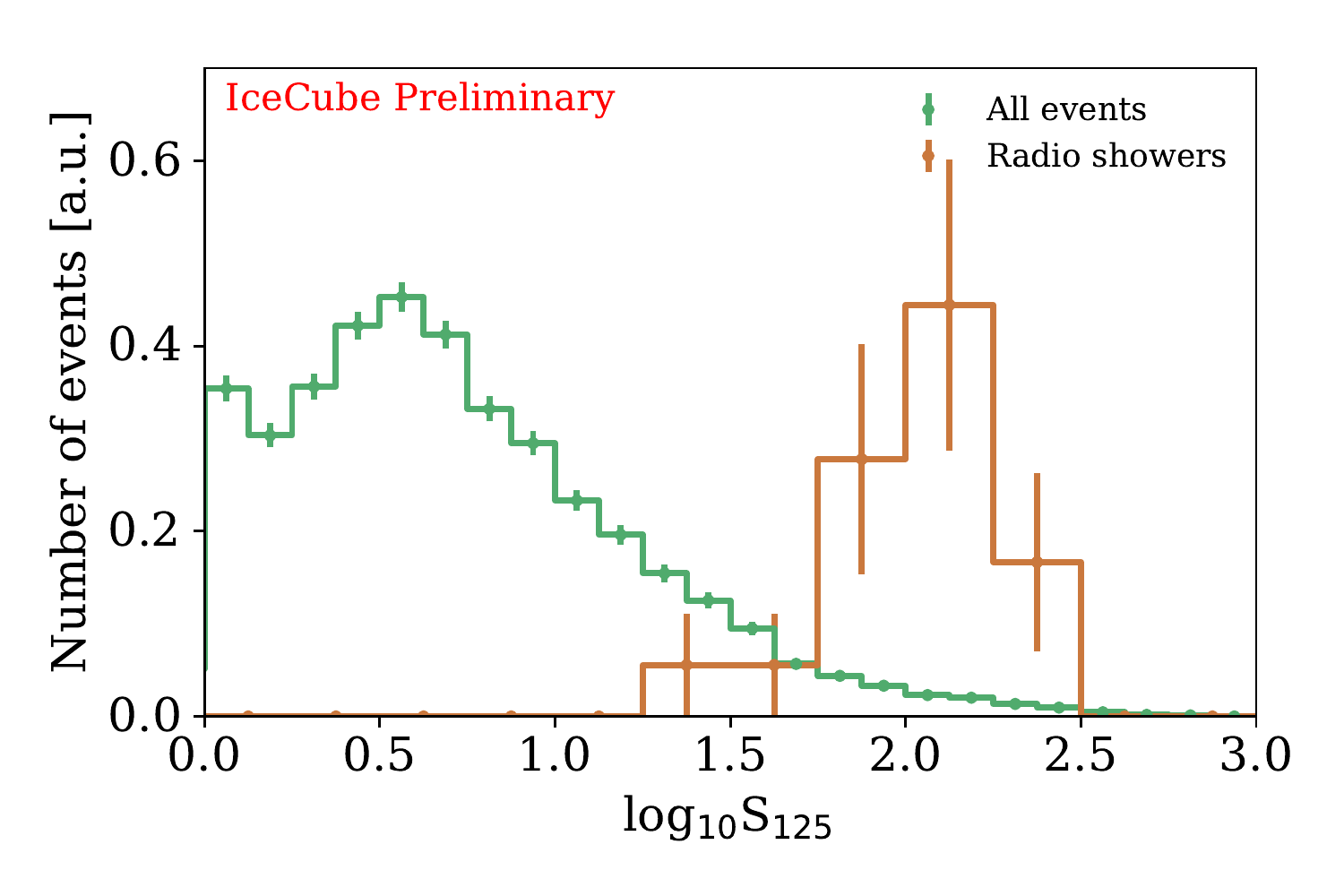}
	\protect\caption{S$_{125}$ distribution}
    \label{fig:s125_dists}
\end{subfigure}
\caption{Distributions of all the coincidence events (green) and of events that pass the radio cuts (red). The error bars represent statistical 1~$\sigma$ uncertainties. The S$_{125}$ is an energy-proxy parameter reconstructed by IceTop and is given in units of VEM. For more detail see~\cite{IT_energy_reco}.}
\label{fig:event_dists}
\end{figure}

The reconstructed zenith direction of all the showers and the showers passing the radio cuts are compared in Figure~\ref{fig:zenith_dist}. A few interesting observations can be made here. Low zenith (vertical) events generally do not pass the radio selection cuts. This is consistent with expectations, as the magnetic field at the South Pole is nearly vertical and the radio emissions from showers parallel to the magnetic field are minimal~\cite{alan2021icrc}. In general the fraction of events passing the radio cuts seems to increase with the increasing zenith, although the statistics are too low at this point to precisely quantify this effect. For zenith angles above \textasciitilde 40°, a gradual fall-off of event rates in all events can be seen. This can be understood as a geometric effect. Since the scintillator panels are fairly two dimensional detectors, their effective area decreases with \textasciitilde cos$\theta$. Combined with the increasing attenuation with slant depth, the trigger threshold occurs at higher energies for larger zenith angles. For vertical showers we also see a decrease in the scintillator trigger rate. One reason for this is because at the moment we have a quite high single-panel trigger threshold (usually >1~MIP), vertical muons are much less likely to lead to a trigger than more inclined muons, which have a longer path length through the detector. The threshold will be reduced down to 0.5 MIP in the future and we expect the issue to go away with that change.

In Figure~\ref{fig:s125_dists}, one can see the distribution of the IceTop-reconstructed energy-proxy parameter, S$_{125}$. The relation of S$_{125}$ to energy can be found in \cite{IT_energy_reco}. However, note that the relation of S$_{125}$ to energy has only ever been calibrated for events with zenith angles up to 37°. Since the events we are looking at here tend to have a zenith angle beyond that, the energy of the events is somewhat uncertain.
Nevertheless, some trends can still be observed. For events passing the radio selection cuts, we can see that their fraction significantly increases with energy. This is also expected behaviour, as the radio detection threshold is significantly higher than that of the scintillators~\cite{agn2019icrc, alan2021icrc}. The lowest S$_{125}$ value observed for a radio shower is 48 VEM (vertical equivalent muons), which would correspond to an energy of around 30-40\,PeV, near the expected energy threshold for detection of the radio  with the IceTop-enhancement~\cite{alan2021icrc}.

\section{Conclusions}\label{sec:conclusions}
A South Pole prototype station of the future IceTop-enhancement has been taking various calibration and air-shower data over the past year. In this work, we had a first look at the cosmic-rays measured with the prototype station. A method for detector calibration and event building has been developed. Basic event reconstruction is performed on the scintillator, radio and IceTop data obtained from the same air-showers. Overall the shower distributions look as expected.

In the future, automated SiPM gain corrections will make the operations more stable. Improved selection cuts will lower the threshold for the detection of radio showers. Improved reconstruction methods and a better combination of the information from the three different detector components should also give us a better angular resolution as well as estimates of the shower energy potentially other shower parameters.

\bibliographystyle{ICRC}
\bibliography{references}
\scriptsize
\smallskip

We acknowledge the support by the Doctoral School "Karlsruhe School of Elementary and Astroparticle Physics: Science and Technology". Also, this project has received funding from the European Research Council (ERC) under the European Union's Horizon 2020 research and innovation programme (grant agreement No 802729).

\clearpage
\section*{Full Author List: IceCube Collaboration}




\scriptsize
\noindent
R. Abbasi$^{17}$,
M. Ackermann$^{59}$,
J. Adams$^{18}$,
J. A. Aguilar$^{12}$,
M. Ahlers$^{22}$,
M. Ahrens$^{50}$,
C. Alispach$^{28}$,
A. A. Alves Jr.$^{31}$,
N. M. Amin$^{42}$,
R. An$^{14}$
% ...
\\

\noindent
% ...
$^{12}$ Universit{\'e} Libre de Bruxelles, Science Faculty CP230, B-1050 Brussels, Belgium \\
% ...
$^{14}$ Department of Physics and Laboratory for Particle Physics and Cosmology, Harvard University, Cambridge, MA 02138, USA \\
% ...
$^{17}$ Department of Physics, Loyola University Chicago, Chicago, IL 60660, USA \\
$^{18}$ Dept. of Physics and Astronomy, University of Canterbury, Private Bag 4800, Christchurch, New Zealand \\
% ...
$^{22}$ Niels Bohr Institute, University of Copenhagen, DK-2100 Copenhagen, Denmark \\
% ...
$^{28}$ D{\'e}partement de physique nucl{\'e}aire et corpusculaire, Universit{\'e} de Gen{\`e}ve, CH-1211 Gen{\`e}ve, Switzerland \\
% ...
$^{31}$ Karlsruhe Institute of Technology, Institute for Astroparticle Physics, D-76021 Karlsruhe, Germany  \\
% ...
$^{42}$ Bartol Research Institute and Dept. of Physics and Astronomy, University of Delaware, Newark, DE 19716, USA \\
% ...
$^{50}$ Oskar Klein Centre and Dept. of Physics, Stockholm University, SE-10691 Stockholm, Sweden \\
% ...
$^{59}$ DESY, D-15738 Zeuthen, Germany \\
% ...

\subsection*{Acknowledgements}

\noindent
USA {\textendash} U.S. National Science Foundation-Office of Polar Programs,
U.S. National Science Foundation-Physics Division,
U.S. National Science Foundation-EPSCoR,
Wisconsin Alumni Research Foundation,
Center for High Throughput Computing (CHTC) at the University of Wisconsin{\textendash}Madison,
Open Science Grid (OSG),
Extreme Science and Engineering Discovery Environment (XSEDE),
Frontera computing project at the Texas Advanced Computing Center,
U.S. Department of Energy-National Energy Research Scientific Computing Center,
Particle astrophysics research computing center at the University of Maryland,
Institute for Cyber-Enabled Research at Michigan State University,
and Astroparticle physics computational facility at Marquette University;
Belgium {\textendash} Funds for Scientific Research (FRS-FNRS and FWO),
FWO Odysseus and Big Science programmes,
and Belgian Federal Science Policy Office (Belspo);
Germany {\textendash} Bundesministerium f{\"u}r Bildung und Forschung (BMBF),
Deutsche Forschungsgemeinschaft (DFG),
Helmholtz Alliance for Astroparticle Physics (HAP),
Initiative and Networking Fund of the Helmholtz Association,
Deutsches Elektronen Synchrotron (DESY),
and High Performance Computing cluster of the RWTH Aachen;
Sweden {\textendash} Swedish Research Council,
Swedish Polar Research Secretariat,
Swedish National Infrastructure for Computing (SNIC),
and Knut and Alice Wallenberg Foundation;
Australia {\textendash} Australian Research Council;
Canada {\textendash} Natural Sciences and Engineering Research Council of Canada,
Calcul Qu{\'e}bec, Compute Ontario, Canada Foundation for Innovation, WestGrid, and Compute Canada;
Denmark {\textendash} Villum Fonden and Carlsberg Foundation;
New Zealand {\textendash} Marsden Fund;
Japan {\textendash} Japan Society for Promotion of Science (JSPS)
and Institute for Global Prominent Research (IGPR) of Chiba University;
Korea {\textendash} National Research Foundation of Korea (NRF);
Switzerland {\textendash} Swiss National Science Foundation (SNSF);
United Kingdom {\textendash} Department of Physics, University of Oxford.

\end{document}